\documentclass[twocolumn]{aastex631}

\newcommand{\eg}{e.g., }
\newcommand{\ie}{i.e., }

\shorttitle{Cerium features in kilonova near-infrared spectra}
\shortauthors{Tanaka, Domoto, Aoki et al.}
\graphicspath{{./}{figures/}}

\begin{document}

\title{Cerium features in kilonova near-infrared spectra: implication from a chemically peculiar star}

\author[0000-0001-8253-6850]{Masaomi Tanaka}
\affiliation{Astronomical Institute, Tohoku University, Aoba, Sendai 980-8578, Japan}
\affiliation{Division for the Establishment of Frontier Sciences, Organization for Advanced Studies, Tohoku University, Sendai 980-8577, Japan}

\author[0000-0002-7415-7954]{Nanae Domoto}
\affiliation{Astronomical Institute, Tohoku University, Aoba, Sendai 980-8578, Japan}

\author[0000-0002-8975-6829]{Wako Aoki}
\affiliation{National Astronomical Observatory of Japan, 2-21-1 Osawa, Mitaka, Tokyo 181-8588, Japan}
\affiliation{Astronomical Science Program, The Graduate University for Advanced Studies, SOKENDAI, 2-21-1 Osawa, Mitaka, Tokyo 181-8588, Japan}

\author[0000-0003-4656-0241]{Miho N. Ishigaki}
\affiliation{National Astronomical Observatory of Japan, 2-21-1 Osawa, Mitaka, Tokyo 181-8588, Japan}

\author[0000-0002-4759-7794]{Shinya Wanajo}
\affiliation{Max-Planck-Institut f\"ur Gravitationsphysik (Albert-Einstein-Institut), Am M\"uhlenberg 1, D-14476 Potsdam-Golm, Germany}

\author[0000-0002-2502-3730]{Kenta Hotokezaka}
\affiliation{Research Center for the Early Universe, Graduate School of Science, University of Tokyo, Bunkyo, Tokyo 113-0033, Japan}
\affiliation{Kavli IPMU (WPI), UTIAS, The University of Tokyo, Kashiwa, Chiba 277-8583, Japan}

\author[0000-0003-4443-6984]{Kyohei Kawaguchi}
\affiliation{Max-Planck-Institut f\"ur Gravitationsphysik (Albert-Einstein-Institut), Am M\"uhlenberg 1, D-14476 Potsdam-Golm, Germany}
\affiliation{Institute for Cosmic Ray Research, The University of Tokyo, Kashiwa, Chiba 277-8582, Japan}
\affiliation{Center for Gravitational Physics, Yukawa Institute for Theoretical Physics, Kyoto University, Kyoto 606-8502, Japan}

\author[0000-0002-5302-073X]{Daiji Kato}
\affiliation{National Institute for Fusion Science, 322-6 Oroshi-cho, Toki 509-5292, Japan}
\affiliation{Interdisciplinary Graduate School of Engineering Sciences, Kyushu University, Kasuga, Fukuoka 816-8580, Japan}

\author[0000-0003-0894-7824]{Jae-Joon Lee}
\affiliation{Korea Astronomy and Space Science Institute, 776 Daedeok-daero, Yuseong-gu, Daejeon 34055, Republic of Korea}
  
\author[0000-0002-3808-7143]{Ho-Gyu Lee}
\affiliation{Korea Astronomy and Space Science Institute, 776 Daedeok-daero, Yuseong-gu, Daejeon 34055, Republic of Korea}
\affiliation{Space Policy Research Center, Science and Technology Policy Institute, 370 Sicheong-daero, Sejong, 30147, Republic of Korea}

\author[0000-0003-3618-7535]{Teruyuki Hirano}
\affiliation{Astrobiology Center, 2-21-1 Osawa, Mitaka, Tokyo 181-8588, Japan}
\affiliation{National Astronomical Observatory of Japan, 2-21-1 Osawa, Mitaka, Tokyo 181-8588, Japan}
\affiliation{Astronomical Science Program, The Graduate University for Advanced Studies, SOKENDAI, 2-21-1 Osawa, Mitaka, Tokyo 181-8588, Japan}

\author[0000-0001-6181-3142]{Takayuki Kotani}
\affiliation{Astrobiology Center, 2-21-1 Osawa, Mitaka, Tokyo 181-8588, Japan}
\affiliation{National Astronomical Observatory of Japan, 2-21-1 Osawa, Mitaka, Tokyo 181-8588, Japan}
\affiliation{Astronomical Science Program, The Graduate University for Advanced Studies, SOKENDAI, 2-21-1 Osawa, Mitaka, Tokyo 181-8588, Japan}

\author[0000-0002-4677-9182]{Masayuki Kuzuhara}
\affiliation{Astrobiology Center, 2-21-1 Osawa, Mitaka, Tokyo 181-8588, Japan}
\affiliation{National Astronomical Observatory of Japan, 2-21-1 Osawa, Mitaka, Tokyo 181-8588, Japan}

\author[0000-0001-9326-8134]{Jun Nishikawa}
\affiliation{Astrobiology Center, 2-21-1 Osawa, Mitaka, Tokyo 181-8588, Japan}
\affiliation{National Astronomical Observatory of Japan, 2-21-1 Osawa, Mitaka, Tokyo 181-8588, Japan}
\affiliation{Astronomical Science Program, The Graduate University for Advanced Studies, SOKENDAI, 2-21-1 Osawa, Mitaka, Tokyo 181-8588, Japan}

\author[0000-0002-5051-6027]{Masashi Omiya}
\affiliation{Astrobiology Center, 2-21-1 Osawa, Mitaka, Tokyo 181-8588, Japan}
\affiliation{National Astronomical Observatory of Japan, 2-21-1 Osawa, Mitaka, Tokyo 181-8588, Japan}

\author[0000-0002-6510-0681]{Motohide Tamura}
\affiliation{Astrobiology Center, 2-21-1 Osawa, Mitaka, Tokyo 181-8588, Japan}
\affiliation{National Astronomical Observatory of Japan, 2-21-1 Osawa, Mitaka, Tokyo 181-8588, Japan}
\affiliation{Department of Astronomy, Graduate School of Science, The University of Tokyo, 7-3-1 Hongo, Bunkyo-ku, Tokyo 113-0033, Japan}

\author{Akitoshi Ueda}
\affiliation{Astrobiology Center, 2-21-1 Osawa, Mitaka, Tokyo 181-8588, Japan}
\affiliation{National Astronomical Observatory of Japan, 2-21-1 Osawa, Mitaka, Tokyo 181-8588, Japan}
\affiliation{Astronomical Science Program, The Graduate University for Advanced Studies, SOKENDAI, 2-21-1 Osawa, Mitaka, Tokyo 181-8588, Japan}



\begin{abstract}
  Observations of the kilonova from a neutron star merger event GW170817
  opened a way to directly study $r$-process nucleosynthesis
  by neutron star mergers.
  It is, however, challenging to identify the individual elements
  in the kilonova spectra due to lack of complete atomic data,
  in particular, at near-infrared wavelengths.
  In this paper, we demonstrate that spectra of chemically peculiar
  stars with enhanced heavy element abundances can provide us with
  an excellent astrophysical laboratory for kilonova spectra.
  We show that the photosphere of a late B-type chemically peculiar star
  HR 465 has similar lanthanide abundances and ionization degrees with those
  in the line forming region in a kilonova at $\sim 2.5$ days after the merger.
  The near-infrared spectrum of HR 465 taken with Subaru/IRD
  indicates that Ce III lines give the strongest absorption features
  around 16,000 \AA\ and
  there are no other comparably strong transitions around these lines.  
  The Ce III lines nicely match with the broad absorption features at 14,500 \AA\
  observed in GW170817 with a blueshift of $v=0.1c$,
  which supports recent identification
  of this feature as Ce III by \citet{domoto22}.
\end{abstract}



\section{Introduction} \label{sec:intro}

Multi-messenger observations of a neutron star merger
GW170817 opened a new pathway to study the astrophysical sites
of $r$-process nucleosynthesis.
Following the detection of gravitational waves,
an electromagnetic counterpart was identified in
ultraviolet, optical, and infrared wavelengths (AT 2017gfo,
\citealt{abbott17MMA}).
The properties of AT 2017gfo are consistent
with those of a kilonova \citep[\eg][]{li98,metzger10},
a thermal emission powered by decays of $r$-process nuclei,
which gives evidence that 
the neutron star merger is a site of $r$-process nucleosynthesis
\citep[\eg][]{kasen17,perego17,tanaka17,rosswog18,kawaguchi18}.

To extract detailed information of $r$-process nucleosynthesis,
such as the amounts of individual elements and their spatial distribution,
it is necessary to identify elements in the spectra
as done for stellar spectra.
In fact, a series of optical and near-infrared (NIR)
spectra have been taken for AT 2017gfo
\citep[\eg][]{chornock17,pian17,smartt17}.
However, it is challenging to decode the spectra
due to high Doppler shifts of the spectral lines,
presence of many heavy elements,
and lack of complete and accurate atomic data,
in particular, in the NIR wavelengths.

\citet{watson19} proposed that the observed absorption
features around 9,000 \AA\ are caused by Sr II lines.
This identification was further confirmed by
\citet{domoto21} with full radiative transfer simulations.
Also, \citet{gillanders22} quantified the Sr abundance
by atmospheric modeling.
The same feature may also be caused by a He I line
\citep{perego22}, although the strength of the He I line
depends on non local thermodynamic equilibrium (non-LTE) effects
\citep{tarumi23}.

Recently, \citet{domoto22} systematically studied
the spectral features of kilonovae in the NIR wavelengths.
They identified several elements producing strong transitions
in kilonovae, such as Ca, Sr, Y, Zr, Ba, La, and Ce.
Then, they constructed a hybrid linelist by combining
theoretical linelist based on atomic models of \citet{tanaka20}
and accurate linelist for the important ions
based on spectroscopic experiments.
By performing radiative transfer simulations with the hybrid linelist,
they showed that the broad NIR features around 13,000 \AA\ and 14,500 \AA\
in AT 2017gfo are reproduced by La III and Ce III lines, respectively.

There are, however, still no {\it complete} and {\it accurate}
atomic data at NIR wavelengths,
which cover all the heavy elements synthesized in the neutron star mergers.
Therefore, there might be unknown transitions of other elements
that contribute to the NIR absorption features of kilonova spectra.

In this paper, we demonstrate that
atmosphere of chemically peculiar stars 
provides an excellent laboratory to decode kilonova spectra.
Chemically peculiar stars are known to exhibit abnormal elemental abundance patterns.
Some chemically peculiar stars show extremely enhanced metal abundances
(hereafter we call stars with enhanced metal abundances as
chemically peculiar stars),
which are likely to be caused by atomic diffusion \citep{michaud70}.
Thus, the atmosphere of such stars can mimic the elemental abundances
in kilonovae dominated by heavy elements.
In Section \ref{sec:cpstar}, we show similarities
in the atmospheres of chemically peculiar stars
and kilonovae by analyzing the properties of
a late B-type chemically peculiar star HR 465
(= HD 9996, e.g., \citealt{preston70,aller72,cowley87}).
In Section \ref{sec:hr465}, we show the NIR spectrum
of HR 465 and measurements of strong absorption lines.
Then, we discuss implications to kilonova spectra in Section \ref{sec:kilonova}.
Finally we give concluding remarks in Section \ref{sec:conclusions}.
All the wavelengths given in this paper are those in air.


\begin{figure}
\begin{center}
\includegraphics[width=8.cm]{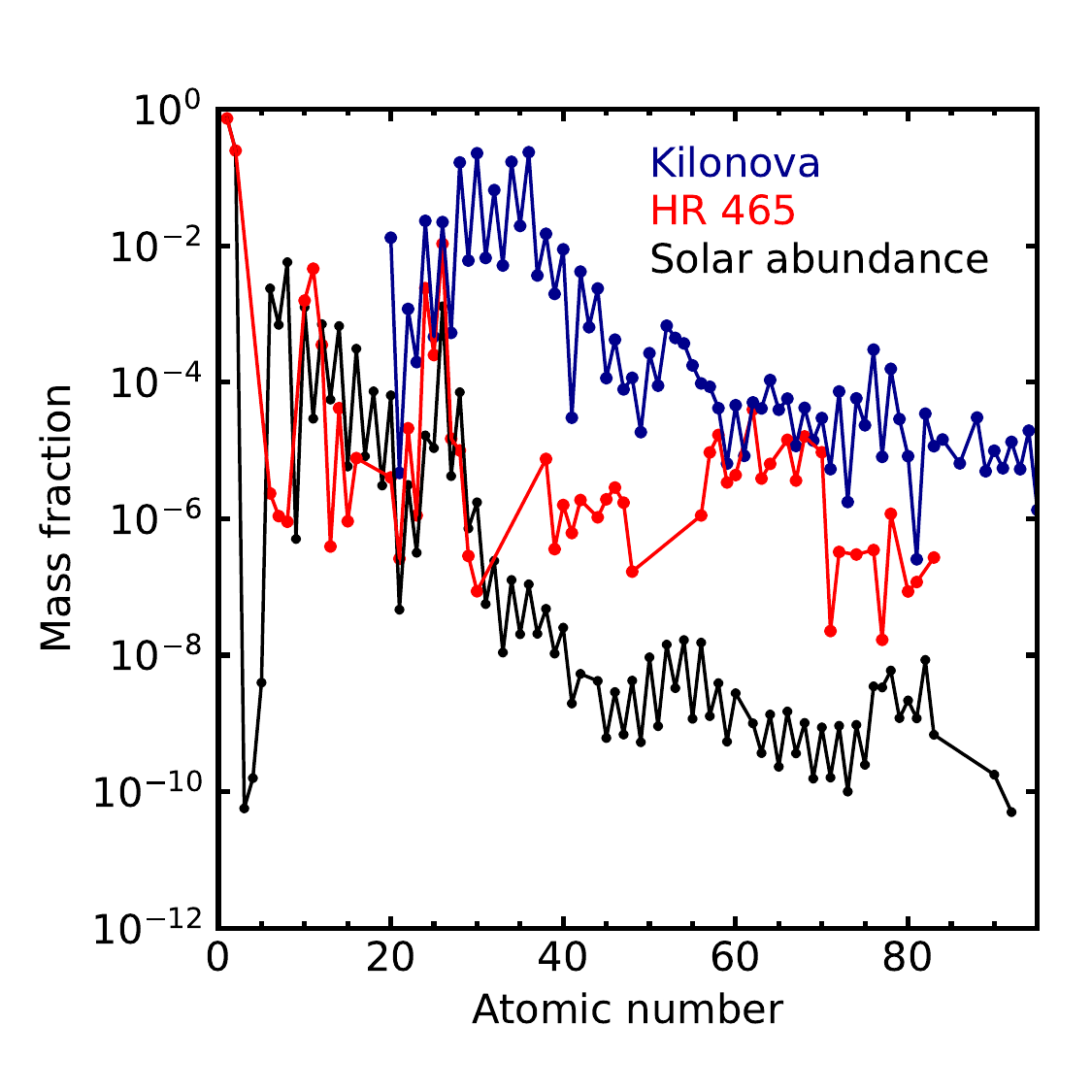}\\
\includegraphics[width=8.cm]{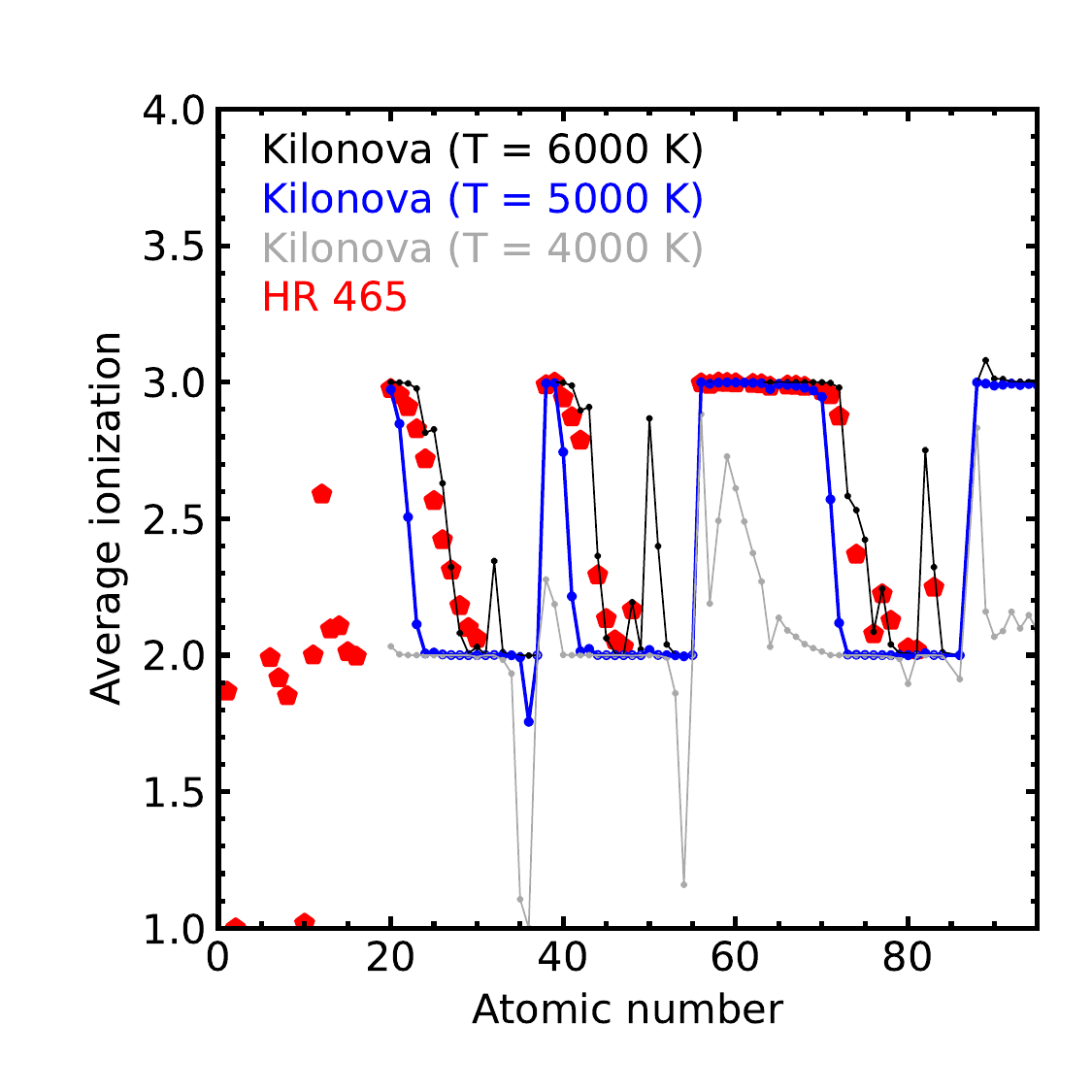}\\  
\caption{{\it Top}: Abundances in kilonova ejecta at $t=2.5$ days
  (blue, \citealt{domoto21,domoto22})
  compared with those of a chemically peculiar star HR 465
  (red, \citealt{nielsen20}) and solar abundances (black).
 {\it Bottom}: Average ionization degrees in the photosphere of kilonovae (black, blue, and gray lines) and HR 465 (red) calculated under the assumption of LTE.
}
\label{fig:abun}
\end{center}
\end{figure}

\begin{figure*}
  \begin{center}
    \begin{tabular}{cc}
  \includegraphics[width=9cm]{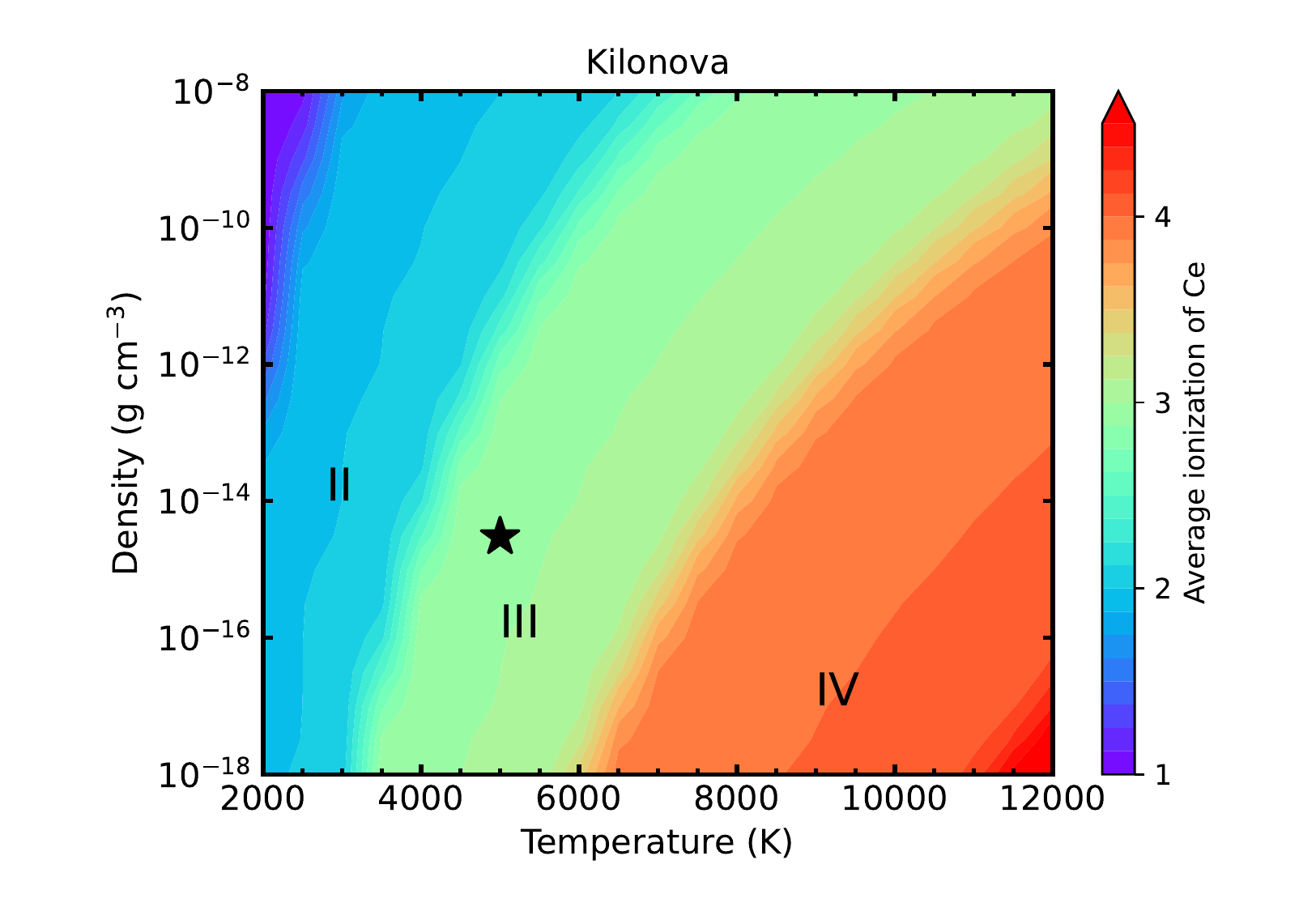}&
  \includegraphics[width=9cm]{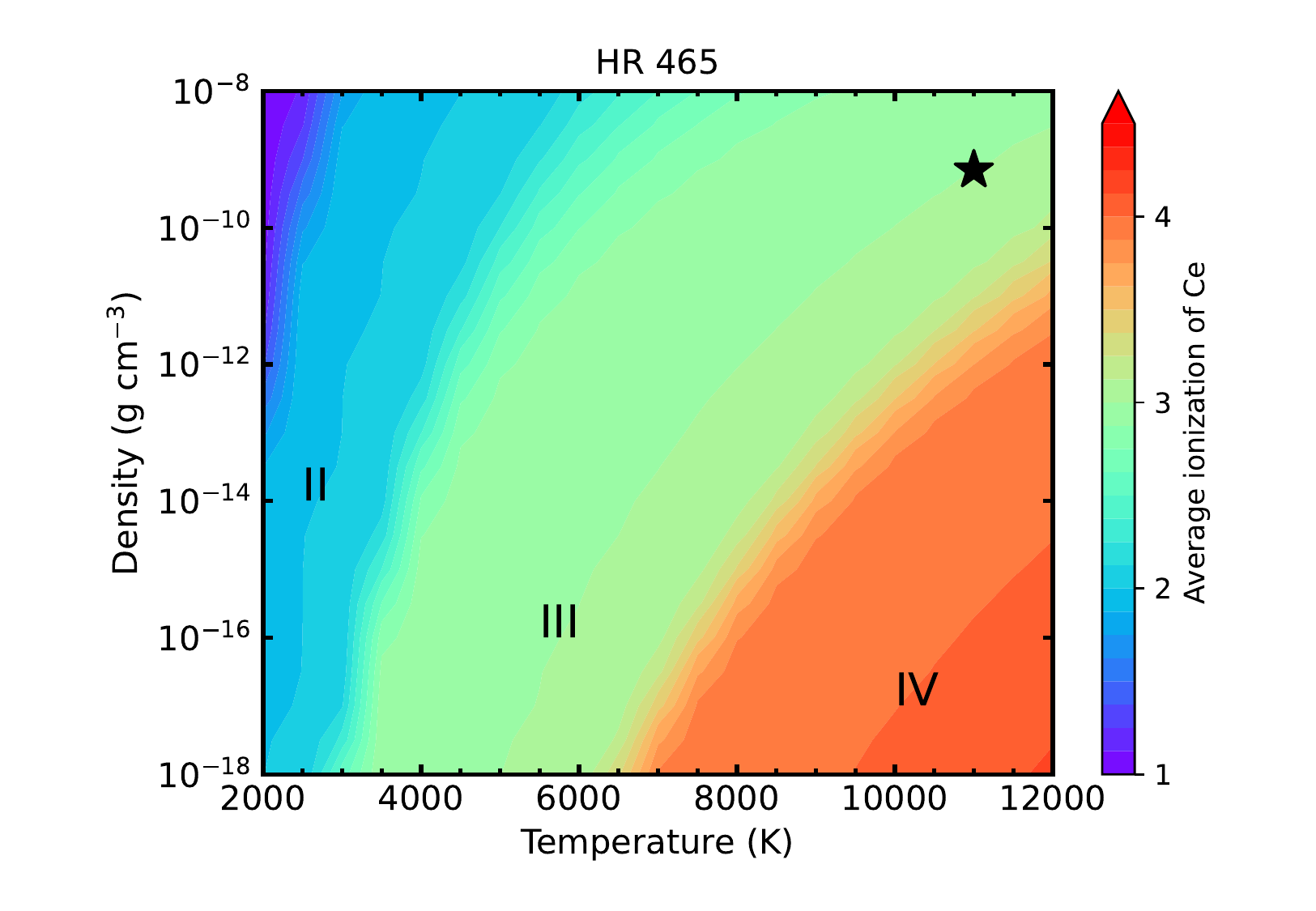}
    \end{tabular}
  \caption{(Left) Average ionization degree of Ce as a function of temperature and density for the abundances of kilonovae (blue points in the top panel of Figure \ref{fig:abun}). (Right) The same for the abundances of HR 465 (red points in the top panel of Figure \ref{fig:abun}). The fiducial parameters for the kilonova ejecta and the photosphere of HR 465 are marked with star symbols. The boundary between doubly (III) and triply (IV) ionized region is different between two cases. This is because of the H-rich abundances in HR 465: the free electrons are mainly provided by ionized H at a high temperature.
}
\label{fig:ion}
\end{center}
\end{figure*}

\begin{figure*}
\begin{center}
\includegraphics[width=16cm]{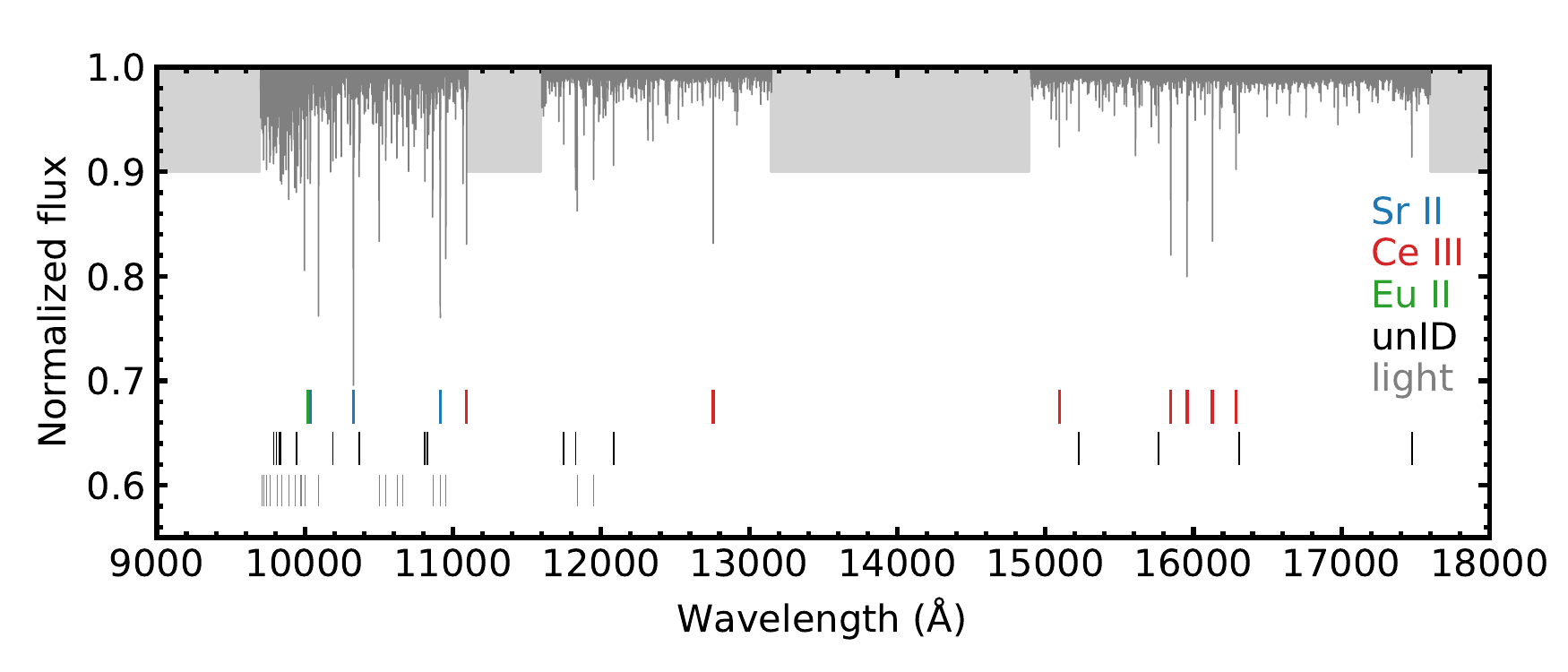}\\
\includegraphics[width=18cm]{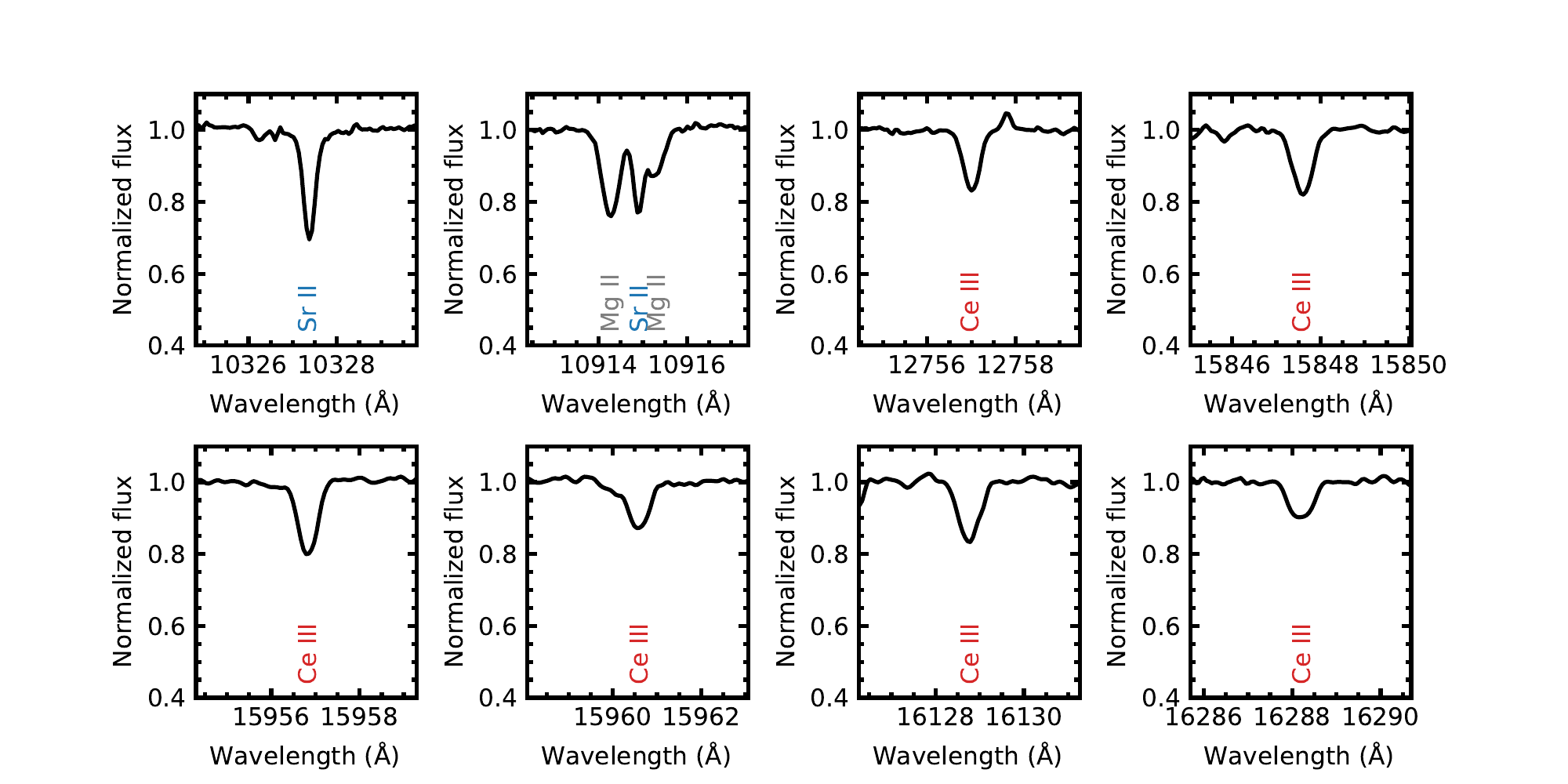}
\caption{Normalized NIR spectrum of HR 465 (top) and that around 8 strongest transitions of heavy elements (bottom). In the top panel, the strongest 50 lines are indicated.
The lines of Sr II, Ce III, and Eu II are shown in red, blue, and green, respectively.
The lines of the elements lighter than Fe are shown in gray. The lines shown in black are not matched with the linelists (unID).
The gray shaded regions show the wavelength range
with strong atmospheric absorption.
}
\label{fig:spec_zoom}
\end{center}
\end{figure*}

\begin{figure*}
\begin{center}
  \includegraphics[width=12cm]{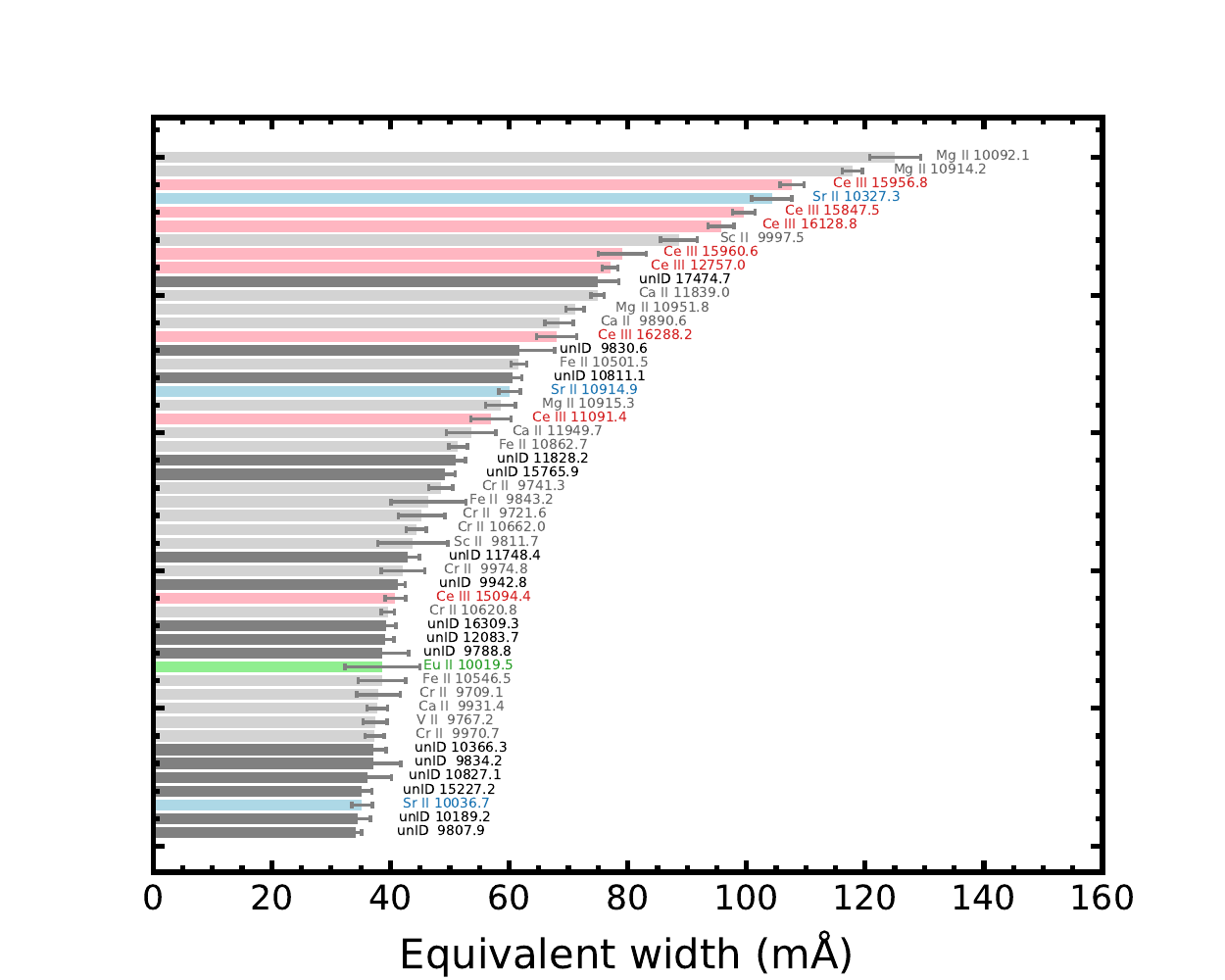}
  \caption{Equivalent width of the 50 strongest lines in the NIR spectrum of HR 465.
    The lines of elements heavier than Fe, which are abundant in kilonova ejecta, are shown in red (Ce), blue (Sr), and green (Eu), while the lines of lighter elements are shown in gray.
    The unID lines that can also potentially appear in kilonova spectra are shown in black.
}
\label{fig:ew}
\end{center}
\end{figure*}

\section{Chemically peculiar star as a laboratory of kilonova spectrum}
\label{sec:cpstar}

\subsection{Abundances}

Abundances in the ejected material from neutron star mergers
(hereafter kilonova ejecta)
are dominated by elements heavier than Fe.
The top panel of Figure \ref{fig:abun} shows an abundance distribution
adopted in the radiative transfer simulations
by \citet{domoto21} and \citet{domoto22} as a representative case
of kilonova ejecta
(their ``light'' model assuming a solar $r$-process-like pattern with an enhanced light component; see also \citealt{wanajo18}).
This particular example is a case that is consistent with
the abundances of metal-poor stars with a weak $r$-process
signature such as HD 122563 \citep{honda06}.

Chemically peculiar stars often show enhanced abundances of elements
heavier than Fe \citep[e.g.,][]{ghazaryan18}.
In particular, Ap and Bp type stars show the abundances of lanthanides
(atomic number $Z=57-71$)
enhanced by more than two orders of magnitude,
which can match the high heavy element abundances in kilonovae.
The red points in Figure \ref{fig:abun} (top panel) show the
elemental abundances of HR 465 (B9p, \citealt{nielsen20}).
HR 465 is an intensively studied chemically peculiar star
\citep[\eg][]{preston70,aller72,cowley87}
with a metallicity of [Fe/H] $\simeq +1.0$ \citep{nielsen20}.
As clearly shown in Figure \ref{fig:abun} (top panel),
the abundances of heavy elements in HR 465
are greatly enhanced as compared with the solar abundance pattern
(black, \citealt{asplund09})
\footnote{The abundances of HR 465 are known to
evolve with a photometric/magnetic phase \citep{rice88}
with a period of 21.5 yr \citep{pyper17}.
The abundances of the rare-earth elements become
  maximum at $\phi \simeq 0$ and minimum at $\phi \simeq 0.5$.
Our IRD observations (Section \ref{sec:hr465}) are performed at $\phi = 0.77$.
We show the abundances in Figure \ref{fig:abun} by taking
average of the abundances
derived from the lines of different ionization degrees (if any)
at different phases \citep[$\phi$ = 0.45, 0.68, and 0.85][]{nielsen20}.
The variation is up to about one order of magnitude,
and the global trend discussed in this paper is not affected.
}.

In particular, the abundances of lanthanides ($Z=57-71$) in HR 465
are enhanced by about 3 orders of magnitude,
and their mass fractions are remarkably similar to those in the kilonova ejecta.
Lanthanides are known to produce strong lines at NIR wavelengths
due to their low-lying, dense energy levels \citep{kasen13,tanaka13,fontes20}.
Thus, thanks to the similarity in the abundances,
NIR spectra of such chemically peculiar stars
can provide the candidates for strong absorption lines of lanthanides
in kilonova NIR spectra.

\subsection{Ionization degrees}

Some chemically peculiar stars also have similar ionization
degrees with those in kilonova ejecta in a few days after the merger.
The bottom panel of Figure \ref{fig:abun} compares ionization degrees
of kilonova and those of HR 465.
The lines show average ionization degrees
(neutral is denoted as 1, single ionization is denoted as 2, and so on)
in the line forming region in a kilonova calculated under the assumption of LTE.
The fiducial case (blue line) is calculated by assuming 
the abundance distribution as in the top panel,
with a temperature of $T = $ 5000 K
and a matter density of $\rho = 3 \times 10^{-15} \ {\rm g \ cm^{-3}}$.
These parameters are chosen based on the
  properties of the line forming region
in the NIR wavelength ($v=0.1c$)
at $t=2.5$ days after the merger in the radiative transfer simulations
(see Figure 6 of \citealt{domoto22}).
Typically, heavy elements are singly or doubly ionized under these conditions,
and most of lanthanides are doubly ionized.

The temperature assumed here is somewhat higher than those
derived by blackbody fitting to the observed kilonova spectra
($T \sim 4000$ K e.g., \citealt{watson19} and \citealt{sneppen23}).
The difference stems from the fact that the line forming region
in the NIR lines is located at an inner layer
($v \sim 0.1c$ rather than the ``photospheric'' velocity of $v \sim 0.25c$)
due to the lower opacities in the NIR wavelengths \citep{tanaka20}.
To see the dependence of the ionization to the temperature,
we also show the ionization degrees at $T = 6000$ K (black line in the bottom panel of Figure \ref{fig:abun})
and $T = 4000 $ K (gray line).
Also, the left panel of Figure \ref{fig:ion} shows
the average ionization of Ce for a wide range of temperature and density.
Under the LTE, ionization degrees in kilonova ejecta depend on the temperature
rather strongly around $T = 5000$ K.
For example, the latter half of lanthanides become singly ionized at $T = 4000$ K
(gray line in the bottom panel of Figure \ref{fig:abun}).
Note that, as we neglect ionization by non-thermal electrons
produced by beta decay of radioactive nuclei,
the ionization degrees tend to be underestimated (see also Section \ref{sec:conclusions}).

The dominance of doubly ionized lanthanides is also seen
in late B-type or early A-type dwarf stars with
an effective temperature of $T_{\rm eff} \sim 10,000$ K.
The red points in the bottom panel of Figure \ref{fig:abun} shows
average ionization degrees at the photosphere of HR 465:
$T=11,000$ K and $\rho = 7 \times 10^{-10}\ {\rm g\ cm^{-3}}$.
These parameters are taken based on the ATLAS9 atmospheric model
\citep{castelli03} with the stellar parameters of HR 465
(log $g = 4.0$ and $T_{\rm eff} = 11,000$ K),
microturbulence velocity of 2 km s$^{-1}$, and metallicity of [M/H] = 0.5,
which is the most metal rich model available\footnote{\url{https://wwwuser.oats.inaf.it/castelli/grids.html}}.
For the calculation of ionization,
we only include the elements with measured abundances.
But this does not affect the ionization degrees
because most of free electrons are provided by H and Fe.

Although the photosphere of HR 465 has a higher temperature
than that of kilonovae, ionization degrees are similar to those in kilonovae.
This behavior is highlighted in Figure \ref{fig:ion}.
For a given temperature, the ionization degrees decrease for a higher density.
Thus, the higher matter density and higher temperature in the stellar
photosphere result in a similar ionization degree with that in kilonovae.
Around the parameter space of HR 465, the ionization degrees
are not very sensitive to the temperature:
doubly ionized lanthanides are still dominant even with $T$ = 11,000 $\pm$ 1000 K.

In summary, the photosphere of late B-type or early A-type dwarf stars
show similar ionization degrees with those of
the line forming region of kilonovae at early phase.
The photosphere of chemically peculiar stars
has a higher temperature than that in kilonovae, and thus,
for a given ion, more lines can become active
in the photosphere of chemically peculiar stars.
Therefore, the spectra of these stars can provide
the {\it complete} and {\it accurate} linelist for strong transitions
of the elements with a similar abundance to kilonovae
(i.e., lanthanides for the case of HR 465).


\begin{figure*}
\begin{center}
\includegraphics[width=15cm]{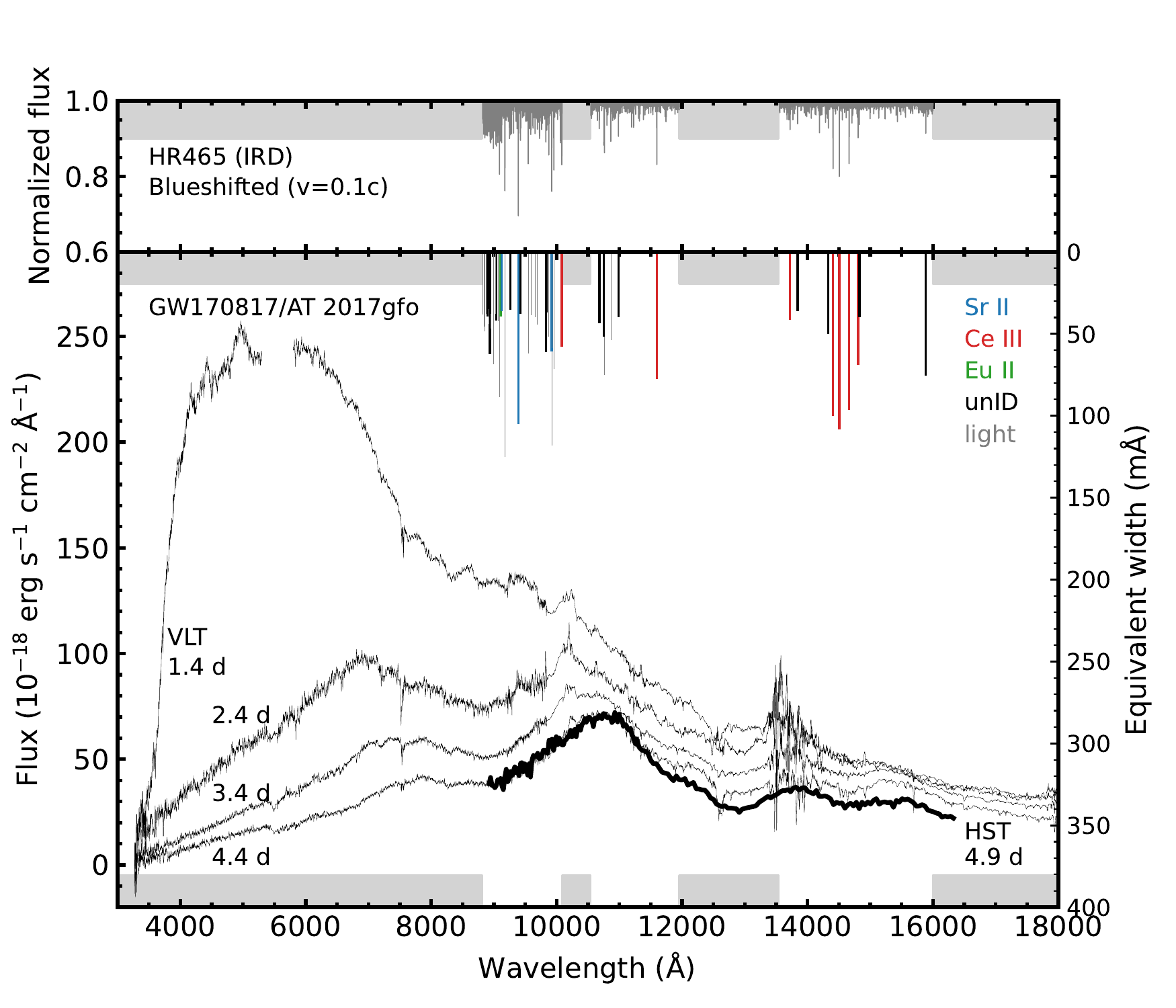}
\caption{(Top) The normalized spectrum of HR 465, blueshifted with $v=0.1c$.
  (Bottom) Spectra of AT 2017gfo taken with
Very Large Telescope (thin lines, \citealt{pian17,smartt17}) and 
Hubble Space Telescope (thick line, \citealt{tanvir17}).
The vertical lines mark the EWs of strong transitions in HR 465,
blueshifted with $v=0.1c$.
The gray shaded region shows the wavelength range
between the atmospheric windows for HR 465 (also blueshifted),
where information of strong transitions are not available.
The colors of EWs are given according to the line identification: Sr II (blue), Ce III (red), Eu II (green), unID (black), and elements lighter than Fe (gray).
}
\label{fig:kilonova}
\end{center}
\end{figure*}

\section{NIR spectrum of HR 465}
\label{sec:hr465}

We obtained the NIR high-resolution spectrum of HR~465 using
the Subaru Telescope InfraRed Doppler instrument
(IRD; \citealt{tamura12}, \citealt{kotani18})
on 2020 July 25 (UT) with an exposure time of 300~sec.
The spectrum covers the wavelength range from 9,300 to 17,600~{\AA}
with a spectral resolution of $R\sim 70000$.
The signal-to-noise ratios of the data range from 150 to 230 per pixel
in the one-dimensional spectra depending
on the position of the blaze profiles of the echelle spectra.
This quality is sufficient for the purpose of the present study.
To our knowledge, this is the first high-resolution
spectrum of HR 465 covering the wide NIR wavelength range.
The details of the observations and data reduction are reported
by \citet{aoki22}.
The telluric absorption lines are correlated by the spectrum
of the bright extremely metal-poor star BD+44$^{\circ}$493,
in which few absorption lines are found in this wavelength range \citep{aoki22}.

The top panel of Figure \ref{fig:spec_zoom} shows the entire spectrum of HR 465
(see Figure \ref{fig:spec1} for the extended view).
The spectrum is normalized by the continuum fitting for the ease of line identification.
Furthermore, the broad hydrogen absorption features are fitted
and the spectrum around those features is flattened.
After masking the wavelength ranges with strong atmospheric absorption,
\ie ranges between $Y$, $J$, and $H$ bands,
as well as artifacts in the telluric correction
(gray shaded regions without line identification in Figure \ref{fig:spec1}),
we detect strong absorption lines.
Then, the detected lines are matched with the VALD linelist \citep{kupka99}
and the NIR linelist by \citet{domoto22}.
There are some lines with which no known transition is matched,
in particular, at shorter wavelengths where more absorption lines exist.
We keep these lines as ``unID'' as they may be caused by heavy elements.
Then, we measure the equivalent widths (EWs) of the strong lines
by assuming Gaussian profile.

Figure \ref{fig:ew} shows the EWs for 50 strongest transitions.
It is clearly seen that, excluding elements lighter than Fe,
the strongest absorptions in the NIR spectrum of HR 465
are dominated by the lines of Ce III and Sr II.
The bottom panels of Figure \ref{fig:spec_zoom} show the spectra around
the six strongest transitions of Ce III and two strongest transitions of Sr II.


\section{Implications to kilonova spectra}
\label{sec:kilonova}

We apply the knowledge from HR 465 to the kilonova spectra.
In Figure \ref{fig:kilonova},
we show the spectrum of HR 465 (top panel) and the EWs of 50 strongest transitions (bottom panel) compared with the spectra of AT 2017gfo
\citep{pian17,smartt17,tanvir17}.
To account for the expansion velocity of the ejecta, 
the lines are blueshifted with a velocity of $v = 0.1$c.
This is a typical velocity of line forming region in the NIR wavelengths
at $t=2.5$ days in the model of \citet{domoto22}.

The absorption features of AT 2017gfo around 14,500 \AA\
  nicely match with the four strong lines of Ce III
  (Ce III $\lambda$15956.8 \AA, $\lambda$15847.5 \AA, $\lambda$16128.8 \AA, and $\lambda$15960.6 \AA)
  with a blueshift of $v = 0.1 c$ (red lines).
Importantly, there are no other elements that cause as
strong contributions as Ce III to this feature.
These facts support the element identification by \citet{domoto22}.

The fact that the temperature of HR 465 is higher than in kilonovae
makes our findings more indicative.
Due to the higher temperature,
the excited states of ions are more populated in HR 465,
and transitions from a higher excitation energy can become active,
which tends to give more candidate ions/transitions
when compared with kilonovae.
However, the lines corresponding to the 14,500 \AA\ feature are
still dominated by the Ce III lines,
and this further supports our identification of the Ce III lines.
Note that the excitation energy of the four strongest transitions of Ce III
  are $E =$ 0.0 eV (ground state), 0.19 eV, 0.39 eV, and 0.81 eV for the Ce III $\lambda$15956.8 \AA, $\lambda$15847.5 \AA, $\lambda$16128.8 \AA, and $\lambda$15960.6 \AA\ lines, respectively.
  Thus, the lower states of these transitions are well populated
  even at the temperature of kilonova ejecta ($T \sim 5000$ K).

It is worth noting that HR 465 also shows the strong Ce III $\lambda$12757.0 \AA\ line, which is a transition from the ground state.
Since this line is as strong as those around 16000 \AA\ and relatively isolated, it may also contribute to the kilonova spectra.
Interestingly, there is a hint of a weak trough around 11500 \AA\ in the HST spectrum of AT 2017gfo \citep{tanvir17}.
If this tentative identification is tested by high-quality, time-series spectra of kilonovae detected in the future,
it strengthens the identification of the Ce III lines
in the NIR wavelength range.

Also, the absorption features around 9,000 \AA\ match with the Sr II lines.
Note that the velocity of the line forming region depends on the
wavelengths due to the wavelength dependence of the opacity:
the line forming region is located at higher velocities
in optical wavelengths due to the higher optical opacity.
A velocity of $v \sim 0.2c$ is more consistent
with the observed feature in AT 2017gfo
as also demonstrated by \citet{watson19}.
As the Sr abundances are quite different between HR 465 and kilonovae
  (top panel of Figure \ref{fig:abun}) and the lower levels of these Sr II lines have a relatively
  high excitation energy ($E=1.8$ eV), the relative line strengths in HR 465
  do not necessarily represent the relative line strengths in kilonovae.

The broad feature at 13,000 \AA\ in the kilonova was identified as
La III by \citet{domoto22}.
However, we could not test this identification with HR 465
because these lines are located in the wavelength range between the atmospheric windows in the rest wavelengths.
To test this identification with stellar spectra,
it is necessary to take high-resolution spectra between $J$ and $H$ bands.

\section{Concluding Remarks}
\label{sec:conclusions}

We demonstrate that spectra of chemically peculiar stars
offer an excellent astrophysical laboratory to decode the kilonova spectra.
Thanks to the enhanced trans-iron abundances and similar ionization degrees,
the absorption features in chemically peculiar stars can
provide the candidates of strong transitions in kilonova spectra.
This is particularly true for lanthanides: their abundances
in chemically peculiar stars can be remarkably similar to those in kilonovae.
By using the NIR spectrum of HR 465, we show that 
(1) the Ce III lines give the dominant contributions to the kilonova feature
around 14,500 \AA\ at early phase, and
(2) there are no other comparably strong transitions in this wavelength range even including unID lines.
These facts support the Ce III identification by \citet{domoto22}.

It should be noted that our analysis assumes LTE for ionization
and excitation.
In a kilonova, non-thermal ionization can be important
in particular at later phases
due to the presence of fast electrons from radioactive decays.
Under such conditions, lanthanides are expected to stay
at singly ionized or doubly ionized states \citep{hotokezaka21,pognan22}.
This may explain the presence of the broad absorption feature
around 14,500 \AA\ even at later phases:
the ionization degrees do not significantly evolve with time,
and the same Ce III lines can keep contributing to the spectra.
To test this possibility, detailed calculations are required
by taking non-thermal ionization into account.
Also, both kilonovae and stellar atmospheres have density and temperature
gradients. In particular, the density structure is shallower
in a kilonova, and different regions may contribute to the spectrum.
Thus, analysis with more realistic density structures is necessary
to fully confirm the similarity between kilonovae and stellar atmospheres.

Another caveat is that the abundance ratios of HR 465 are not identical to
those in the kilonova ejecta.
While the lanthanide fractions are quite similar
as shown in Figure \ref{fig:abun} (top panel),
the elements with $Z=30-50$ are less abundant in HR 465.
Strong NIR transitions are
expected to be dominated by lanthanides (and actinides)
due to their low-lying, dense energy levels \citep{domoto22}.
Thus, we consider that it is less likely that many lighter elements
can contribute to the strong NIR transitions.
Nevertheless, it is worth investigating the NIR transitions for
various elements in actual stars
by obtaining high-resolution NIR spectra for
more chemically peculiar stars with different abundance enhancements
as well as different stellar types (temperature and density).
Such data will provide us with clues to decode kilonova spectra
with different abundances at various epochs.

\begin{acknowledgements}
We thank the referee for valuable comments.
This research was supported by JST FOREST Program (Grant Number JPMJFR212Y),
NIFS Collaborative Research Program (NIFS22KIIF005),
the JSPS Grant-in-Aid for Scientific Research (18H05442, 19H00694, 20H00158, 20H05855, 21H04499, 21H04997), 
and the Grant-in-Aid for JSPS Fellows (22KJ0317).
N.D. acknowledges support from Graduate Program on Physics
for the Universe (GP-PU) at Tohoku University.
\end{acknowledgements}

\appendix

\begin{figure*}
  \begin{center}
  \includegraphics[height=18cm, angle=270]{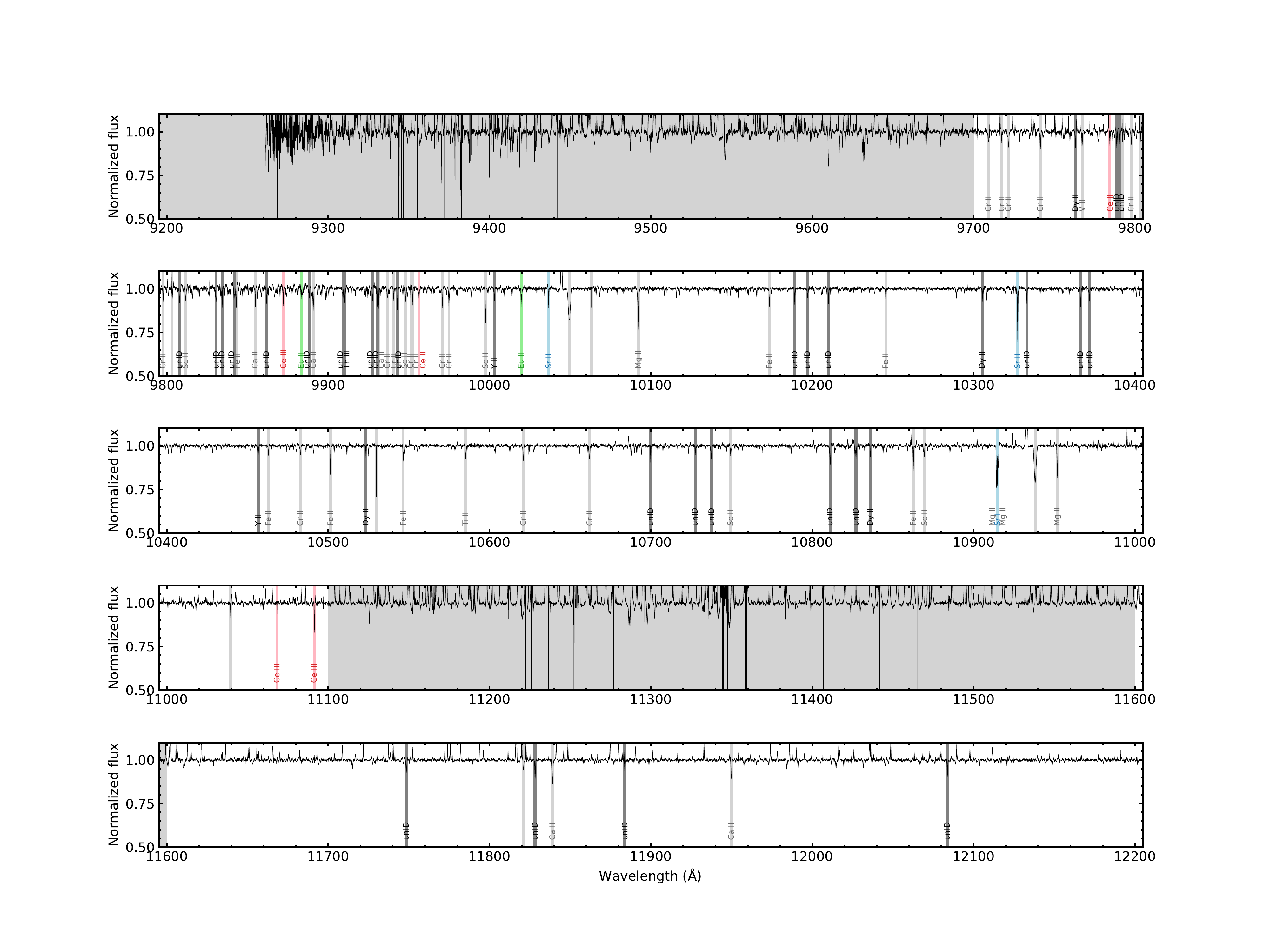}
\caption{Normalized NIR spectrum of HR 465.
}
\label{fig:spec1}
\end{center}
\end{figure*}

Figure \ref{fig:spec1} is an enlarged view of the entire NIR spectrum
of HR 465 taken with Subaru/IRD.
Gray shaded regions without line identification show
the wavelength range with strong telluric absorption and
artifacts in the telluric correction.

\addtocounter{figure}{-1}

\begin{figure*}
  \begin{center}
    \includegraphics[height=18cm, angle=270]{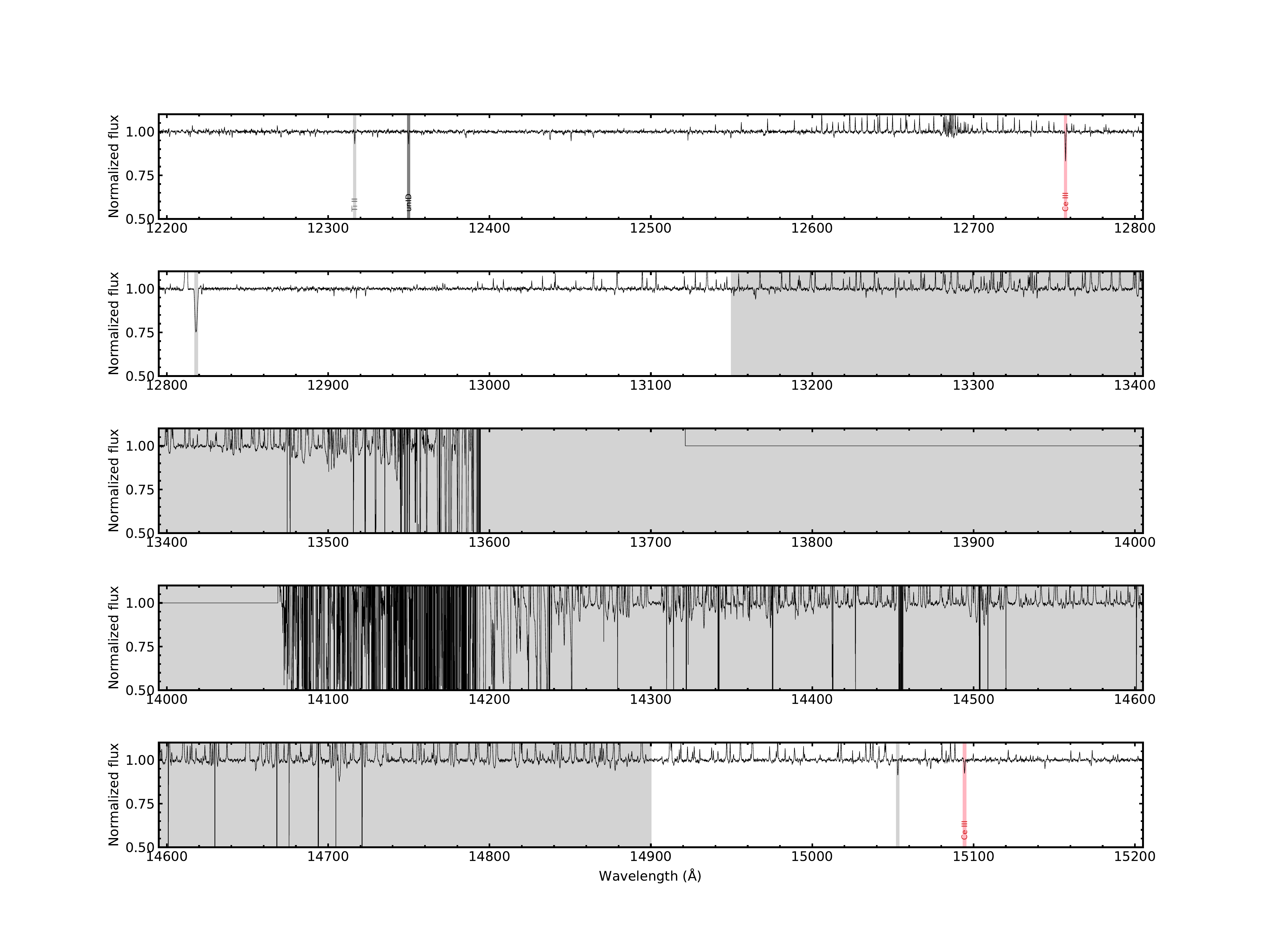}
\caption{Normalized NIR spectrum of HR 465 (cont.)
}
\label{fig:spec2}
\end{center}
\end{figure*}

\addtocounter{figure}{-1}

\begin{figure*}
\begin{center}
    \includegraphics[height=18cm, angle=270]{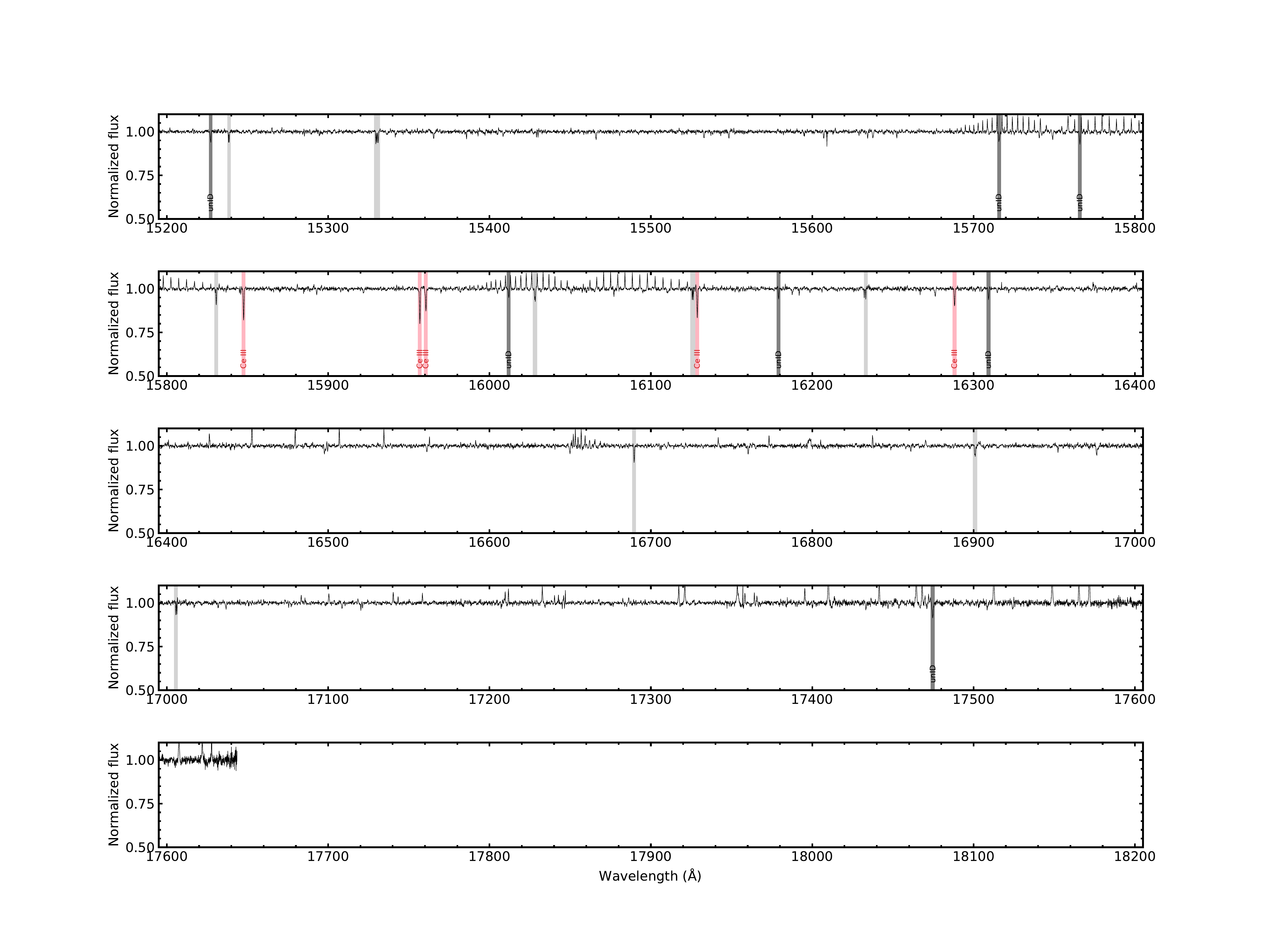}  
\caption{Normalized NIR spectrum of HR 465 (cont.)
}
\label{fig:spec3}
\end{center}
\end{figure*}





\end{document}